\documentclass[]{iopart}

\usepackage{iopams}
\usepackage{epsfig,epsf}
\usepackage{graphicx}

\begin{document}

\title{Baryon enhancement in high-density QCD and relativistic heavy ion collisions}

\author{Yang Li}

\address{Department of Physics and Astronomy, Iowa State University, Ames, IA 50011, USA}
\ead{yangli@iastate.edu}

\begin{abstract}
We argue that the collinear factorization of the fragmentation functions in high energy nuclear collisions breaks down at transverse momenta $p_T \lesssim Q_s/g$ due to high parton densities in the colliding hadrons and/or nuclei. We find that gluon recombination dominates in that $p_T$ region. We calculate the inclusive cross-section for $\pi$ meson and nucleon production using the low energy theorems for the scale anomaly in QCD, and compare our quantitative baryon-to-meson ratio to the RHIC data.
\end{abstract}




\section{Introduction}

It is believed that the strong interactions at high energies/densities are described by the Colour Glass Condensate (CGC), which provides the theoretical framework for the description of the saturation of the gluon fields in the hadronic and nuclear wave functions \cite{CGC}. The transition to the saturation region is marked by the saturation scale $Q_s$, which is the two-dimensional parton colour charge density in the infinite momentum frame of the hadron wave function. The typical momentum scale of the partons in the saturation regime is in the order of $Q_s$.

We argued in \cite{LT1} that the process of gluon fusion in high parton density medium should dominate in the $J^{PC}=0^{++}$ singlet state production channel at $p_T \ll Q_s/\sqrt{\alpha_s}$ region. Contribution of this channel to the total multiplicity is proportional to the spectral density $\rho_\theta(M)$ of the correlator $\langle 0| T \{\theta_\mu^\mu(x),\theta_\nu^\nu(0)\}| 0 \rangle$. In the perturbation theory, valid at large invariant masses, the lowest order contribution to the correlator arises from the two gluon amplitude $|\langle gg|\theta_\mu^\mu(x)|0\rangle|^2$.  On the other hand, at low invariant masses, the main contribution comes from the excitation of hadrons through the $|\langle hh\cdots |\theta_\mu^\mu(x)|0\rangle|^2$ amplitude. We can then apply anomaly matching to infer the hadron production from the two-gluon production amplitude in the same channel by weighting it with an appropriate weight factor, which can be calculated in the chiral perturbation theory.

\section{Gluon pair production in the scale anomaly channel}

\begin{figure}
\centering
\includegraphics[width=6.5cm]{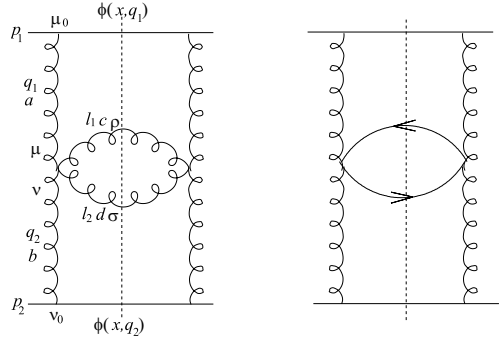}
\caption{Production of gluon and pion pairs in the $J^{PC}=0^{++}$ channel.}
\label{fig:gluon}
\end{figure}

The hadron production model discussed in this paper consists of two stages: (i) the production of a gluon pair in $J^{PC}=0^{++}$ and colour singlet state, and (ii) recombination of this pair into final state hadrons via the anomaly matching mechanism \cite{KharzeevLevin}. It is shown in \cite{LT1} that this gluon fusion process exceeds the collinear factorization channel and dominates in the double inclusive gluon production. The relevant diagrams are given in \fref{fig:gluon}.
In the approximation that gluons are typically produced back-to-back, the spectrum of a single gluon with transverse momentum $p_T$ is given by \cite{KLT}
\begin{equation}\label{eq:y3}
\fl \frac{d\sigma^{0^{++}}_g}{dp_T^2\,dy} = \frac{\pi^4}{16} \int^{4p_T^2} d q_1^2\,\phi_A(x, q_1^2)\,\phi_A(x, 4p_T^2) \int_0^{2Q_s^2} \frac{dM^2}{M^6} \,\frac{\rho_{gg}(M)}{(M^2+4p_T^2)^2} \,,
\end{equation}
where $\phi_A(x,q^2)$ is the unintegrated gluon distribution function and $\rho_{gg}$ is the spectral density in the lowest order perturbation theory \cite{KharzeevLevin}
\begin{equation}\label{eq:spec.den2}
\rho_{gg}(M) = \left(\frac{bg^2}{32\pi^2}\right)^2\frac{2N_c C_F}{4\pi^2}M^4\,.
\end{equation}

\section{Hadronic contributions to the spectral density}

At small invariant masses of the produced gluons, the leading contribution arises from the hadronic excitations and can be calculated in the chiral perturbation theory. The chiral Lagrangian for pions is given by
\begin{equation} \mathcal{L}_\pi = \frac{f_{\pi}^2}{4}{\rm Tr}\,\partial_{\mu}U\,\partial^{\mu}U^{\dagger}+\frac{1}{4}m^2_{\pi}\,f^2_{\pi}\,{\rm Tr}(U+U^{\dagger})\, , \label{eq:PionLagr}
\end{equation}
where $U=\exp\{2i\pi/f_{\pi}\}$, $\pi\equiv \pi^a T^a$ and $T^a$ are the $SU$(2) generators normalized by $\Tr (T^a T^b)=\frac{1}{2} \delta^{ab}$. The non-perturbative contribution for pions to the spectral density can be calculated as $\rho_{\pi\pi}(M^2)=\case{3}{32\pi^2}M^4$, where $M$ is the invariant mass of the initial gluon pair \cite{KharzeevLevin}. For nucleons, the chiral Lagrangian reads
\begin{equation}
\mathcal{L}_N=m_N\bar \psi\, \psi\,, \label{eq:NucleonLagr}
\end{equation}
which gives the spectral density for nucleons $\rho_{N \bar N}(M^2)=\case{4}{32\pi^2}m^2_N M^2$. The spectral density for nucleon kinetic term has a much weaker dependence on the invariant mass $M$, and cannot compete with the $M^4$ dependence for pion kinetic term. Therefore, we need to include the pion-nucleon interaction contribution to the spectral density, which manifest itself as the following Yukawa Lagrangian \cite{WeinbergBook}
\begin{equation}
\mathcal{L}_I = -\frac{2 i m_N g_A}{f_{\pi}}\vec{\pi}\bar{N}\gamma_5\, \vec{t}\, N \, . \label{eq:YukawaLagr}
\end{equation}
The spectral density for pion-nucleon interactions scales as $M^4$ for high invariant masses by evaluating the three-particle final-state diagram \cite{PDB},
\begin{eqnarray}
\rho_ I &=& \left(\frac{m_N g_A}{f_{\pi}}\right)^2 \frac{M^4}{24 \pi^3}\left[\sqrt{1-\frac{4m_N^2}{M^2}}\left(1-16\frac{m_N^2}{M^2}-12\frac{m_N^4}{M^4}\right)\right. \nonumber\\
&-& \left. 72\frac{m_N^4}{M^4}\left(1-\frac{2 m_N^2}{3M^2}\right)\ln \frac{2m_N}{M+\sqrt{M^2 -4m_N^2}}\right] \, , \label{eq:IntRho}
\end{eqnarray}
which ensures that nucleons can be produced with the same orders of magnitude as pions. Their ratio of total cross sections will approach an asymptotic value for high enough colliding energy. Hadron production cross sections can be obtained by replacing the spectral density for double-gluon production in \eref{eq:y3} by the corresponding ones for hadrons in the chiral perturbation theory.

\section{Baryon-to-meson ratio in AuAu collisions at RHIC}

We will now apply our formulas obtained in the previous sections to calculate the $\bar p$ to $\pi^-$ ratio at RHIC energy, which is defined as
\begin{equation}\label{eq:bmr}
R_{\bar p/\pi^-}=\frac{\frac{d\sigma_{\bar p}} {dy\, dp_T^2}  } {\frac{ d\sigma_{\pi^-}}{dy\, dp_T^2}}\approx \frac{\frac{1}{2}\frac{d\sigma_{ N}^{0^{++}}} {dy\, dp_T^2} + \frac{ d\sigma_{\bar p}^{(1)}}{dy\, dp_T^2} } {\frac{1}{3}\frac{ d\sigma_{\pi}^{0^{++}}}{dy\, dp_T^2}+\frac{ d\sigma_{\pi^-}^{(1)}}{dy\, dp_T^2}}\,.
\end{equation}
Here the single hadron production cross section $\frac{d\sigma_{h}^{(1)}}{dy\, dp_T^2}$ is added since it is expected that hard processes will dominate particle productions at high transverse momentum. The coefficients of the soft production channel suggest an isospin symmetry.

The single gluon production cross section decreases as $1/p_T^2$ and will dominate, at some high $p_T$, the one in \eref{eq:y3}, which drops as $1/p_T^6$. The hadron cross section at high $p_T$ is then given by the convolution of gluon production cross section with an appropriate fragmentation function \cite{KKT}
\begin{equation}\label{eq:ktfact2}
\frac{d\sigma_h^{(1)}}{dp_T^2\,dy}=\int \frac{dz}{z^2}\, \frac{d\sigma_g(p_T/z)}{dp_T^2\,dy}\, F^{g/h}_{\rm frag}(z, p_T)\,.
\end{equation}

There is now only one parameter to be specified which is the saturation scale $Q_s$. According to the KLN model \cite{KharzeevLevin, KharzeevNardi} it is given to be the sum of the saturation scales of each nuclei, which at RHIC is $Q^2_s \approx 2\times 0.284\, N_{\rm part}^{1/3}$ GeV$^2$ at midrapidity.

\begin{figure}
\centering
\begin{tabular}{cc}
\includegraphics[width=6cm]{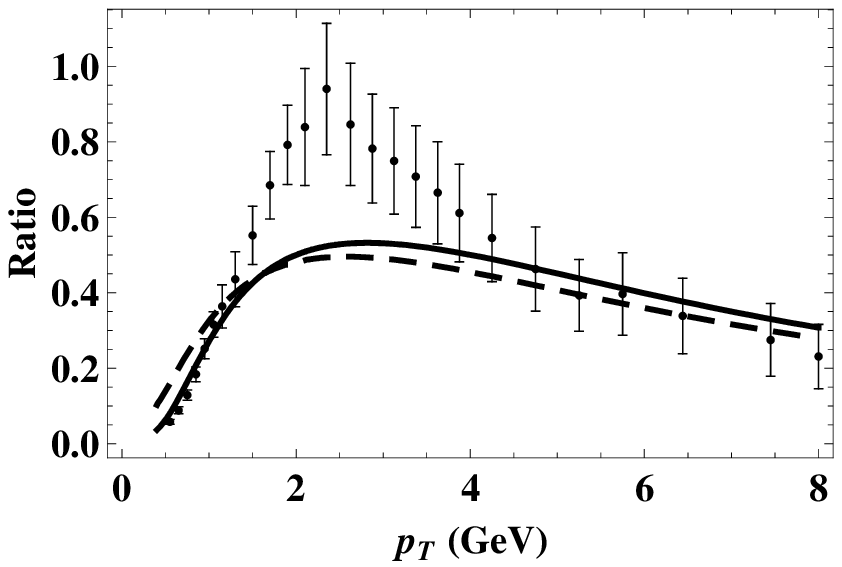} & \includegraphics[width=6cm]{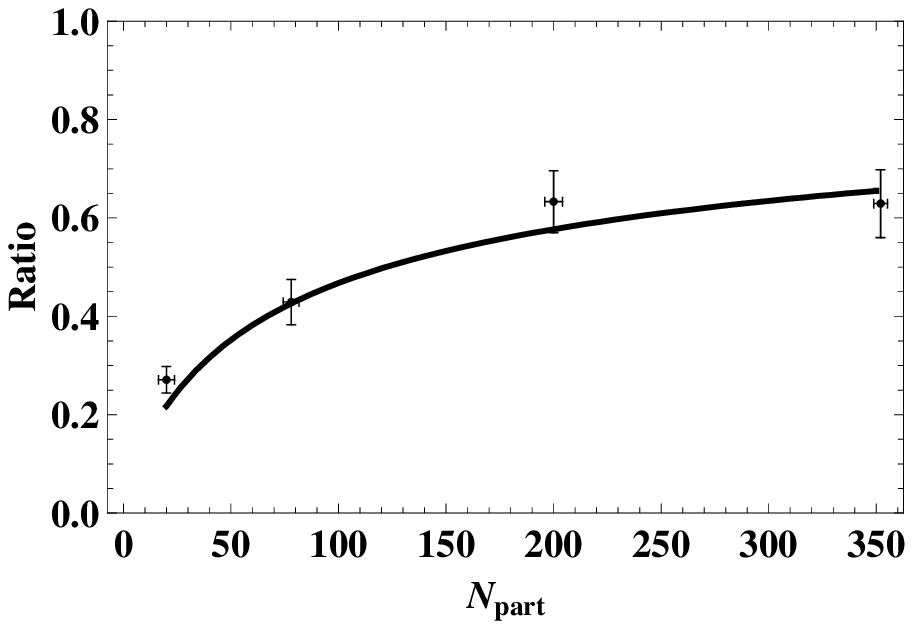}\\
(a) & (b)
\end{tabular}
\caption{Baryon-to-meson ratio in AuAu collisions at $\sqrt{s_{\rm NN}}=200$ GeV as a function of transverse momentum (a) and centrality (b). (a) Data are 0-12\% central AuAu collisions \cite{STARPRL2006}. Two lines correspond to the following relationships between the relative momentum $k$ and the hadron momentum $p_T$. Dashed line: $k^2\approx 4p_T^2$. Solid line: $k^2\approx 4(p_T-\Lambda)^2$, with $\Lambda=0.25$~GeV being a small offset momentum. (b) Data are PHENIX preliminary results for integrated 3 GeV $< p_T<$ 4 GeV region.}
\label{fig:plot}
\end{figure}

The baryon-to-meson ratio (BMR) is plotted as a function of the hadron transverse momentum and the number of participants at $\sqrt{s_{\rm NN}}=200$ GeV in \fref{fig:plot}. We can see the enhancement of the ratio in both cases. Since baryon production has a higher threshold energy, the available phase space is much smaller than for mesons. As a result it is suppressed at small $p_T$ and $N_\mathrm{part}$ compared to meson production. By going from peripheral to central collisions, the saturation scale increases. Hence the increase of a typical gluon transverse momentum makes possible the production of baryons. At $p_T\gg Q_s$ our recombination production mechanism gives way to the traditional gluon fragmentation in accordance with the parton--hadron duality, and the baryon-to-meson ratio is now mainly determined by the pQCD calculations.

Our calculations showed that the cold nuclear effects due to the gluon saturation are reasonably described in AuAu collisions. We believe that this mechanism also contributes to the fast chemical equilibration in the produced dense medium in AuAu collisions by significantly enhancing the baryon yield as compared to the pQCD calculations. We suggest that the ratio is close to the thermal equilibrium value already at the initial time. The final state interaction in the hot medium will achieve a thermal equilibration and may further increase the baryon-to-meson ratio.

\section{Summary}

We have calculated the baryon-to-meson ratio due to cold nuclear effects in the high parton density QCD. We found that a gluon fusion mechanism dominates hadron production and it will significantly enhance the BMR already in the initial time. Combined with pQCD calculation for high transverse momentum region, our model can reasonably describe part of the RHIC AuAu data. Of course, the thermal equilibration is achieved only through the final state interactions.

\ack

I would like to thank my collaborators Kirill Tuchin and Dmitri Kharzeev.

\end{document}